\documentstyle[12pt,aaspp4]{article}
\begin{document}

\title{An 11.6 Micron Keck Search For Exo-Zodiacal Dust}
\author{Marc J. Kuchner, Michael E. Brown and Chris D. Koresko}
\affil{California Institute of Technology, Pasadena, CA 91125}
\authoremail{mjk@gps.caltech.edu, mbrown@gps.caltech.edu, koresko@gps.caltech.edu}

\begin{abstract}

We have begun an observational program to search nearby stars for dust disks
that are analogous to the disk of
zodiacal dust that fills the interior of our solar system.
We imaged six nearby
main-sequence stars with the Keck telescope at 11.6 microns,
correcting for atmosphere-induced wavefront aberrations and
deconvolving the point spread function via classical speckle
analysis.  We compare our data to a simple model of the zodiacal
dust in our own system based on COBE/DIRBE observations (Kelsall et
al. 1998) and place upper limits on the density of exo-zodiacal
dust in these systems.

\end{abstract}


\section {INTRODUCTION}

Our sun is surrounded by a disk of warm ($>$150 K) ``zodiacal'' dust
that radiates most of its thermal energy at 10--30 microns.
This zodiacal dust is produced largely in the inner part of the solar
system by collisions in the asteroid belt
\markcite{derm92}(Dermott et al. 1992) and cometary outgassing \markcite{liouz96}(Liou and Zook 1996).  
Zodiacal dust is interesting as a general feature of planetary systems,
and as an indicator of the presence of larger bodies which supply it;
dust orbiting a few AU from a star is quickly removed as it loses angular
momentum to Poynting-Robertson drag (Robertson 1937).  Understanding
the extra-solar analogs of zodiacal dust may also be crucial in the
search for extra-solar planets \markcite{beich96}
(Beichman et al. 1996) since exo-zodiacal dust in a planetary system
could easily outshine the planets and make them much harder to detect.

The best current upper limits for the existence of
exo-zodiacal dust disks come from IRAS measurements of 12 and 25 micron
excesses above
photospheric emission.  Seen from a nearby star, solar system
zodiacal dust would create only a $10^{-4}$ excess over the sun's
photospheric emission at 20 microns.  IRAS measurements, however, have
typical measurement errors of 5 percent\markcite{mosh92} (Moshir et al. 1992) and display systematic offsets of a similar magnitude when they are compared to other photometry \markcite{cohe96} (Cohen et al. 1996). If there were a solar-type zodiacal disk with 1000 times the density of the disk
around the sun around Tau Ceti, the nearest G star, the excess infrared
emisison would barely exceed the formal 68\% confidence intervals of the IRAS
photometry.  Moreover, all photometric detection schemes of this sort are limited by how accurately the star's mid-infrared photospheric emission is known.   For farther, fainter stars than Tau Ceti, inferring the presence  of dust from the IRAS data becomes still harder.

The detection of faint exo-zodiacal-dust emission is more feasible if one
can resolve the dust emitting region.  The high resolution and dynamic range needed for these observations will generally require large interferometers like the Keck Interferometer, the Large Binocular Telescope, and the Very Large Telescope Interferometer.  But it is already possible to resolve the zodiacal dust mid-infrared emitting regions of the nearest stars.  A 10-meter telescope operating at 12 microns has a diffraction-limited resolution of 0.25 arc seconds, corresponding, for example, to a transverse distance of 2 AU at 8 parsecs.

We have begun a search for zodiacal dust around the nearest stars using the
mid-infrared imaging capabilities of the Long Wavelength Spectrometer (LWS) \markcite{jone93}(Jones \& Peutter 1993) on the W. M. Keck telescope.
The large aperture of the telescope
allows us to make spatially resolved images of the
zodiacal dust 11.6 micron emitting region around the stars so that we
can look for dust emission
above the wings of the point-spread function (PSF) rather than
as a tiny photometric excess against the photosphere.  We present here
the results of two nights of observations, and compare them with a simple model of exo-zodiacal thermal emission to place upper limits on the amount of dust present in the systems we observed.

\section {OBSERVATIONS}

We observed six nearby stars with LWS on the W. M. Keck telescope on
August 3rd and 4th, 1996 using standard mid-infrared imaging
techniques. The target stars were the nearest A--K main-sequence stars
observable from Mauna Kea on those dates.  With the object
on-axis, we took a series of frames lasting 0.8 ms each, chopping the
secondary mirror between the object and blank sky 8 arcseconds to the
north at a frequency of 10 Hz.  Then we nodded the primary
mirror for the next series of frames so that the sky was on-axis and
the object off-axis. We repeated this process for 3 nods over a period of 5
minutes, for an on-source integration time of 1.1 minutes, and a
typical noise of 2 mJy in one 0.11 by 0.11 arcsecond pixel 
due to the thermal background.  The seeing was poor both nights, up to 2 arc seconds in the visible.  To measure the atmosphere-telescope
transfer function, we made similar observations of seven distant,
luminous calibrator stars near our targets on the sky, alternating between target and calibrator every 5--10 minutes.

We increased our frame-rate for the second night of observations so that we could compensate for the seeing using speckle analysis. Figure 1 shows a cut through a single 84 ms exposure of Altair on August 4, compared to an Airy
function representing the diffraction-limited PSF of a filled 10-meter
aperture at 11.6 microns.  The cores of the images are diffraction-limited, but the wings are sensitive to the instantaneous seeing, making speckle analysis necessary. Table 1 provides a summary of our observations.

We flat-fielded the images by comparing the response of each pixel to
the response of a reference pixel near the center of the detector.
First we plotted the data number (DN) recorded by a given pixel
against the DN in the reference pixel for all the frames in each run.
Since the response of each pixel is approximately linear over the 
dynamic range of our observations and most of
the signal is sky background, which varies with
time but is uniform across the chip, the plotted points for each
pixel describe a straight line; if all the pixels had
the same response, the slope of each line would equal 1.  We divided
each pixel's DN by the actual slope of its response curve relative to
the reference pixel, effectively matching all pixels to the reference pixel.
We then interpolated over bad pixels, frame by frame.

To compensate for the differences in the thermal background between the
two nod positions, we averaged together all the on-axis sky frames to
measure the on-axis thermal background and subtracted this average
from all of the on-axis frames---both object and sky.  We used the
same procedure to correct the off-axis frames.

Next, we chose subframes of 32 by 32 pixels on each image, centered
on the star (or for sky frames, the location of the star in an
adjacent object frame), and processed these according to classical
speckle analysis \markcite{labe70}(Labeyrie 1970).  We Fourier
transformed them, and summed the power spectra, yielding a sky
power spectrum and an object power spectrum for each series.  Then
we azimuthally averaged the power spectra in the u-v plane---that
is, we averaged over all the frequency vectors of a given magnitude,
$\sqrt{u^2+v^2}$.  This azimuthal averaging corrects for the rotation
of the focal plane of the alt-az-mounted Keck telescope with respect
to the sky.  We then subtracted from every object power
spectrum the corresponding sky power spectrum and divided each
corrected target power spectrum by the corrected power spectrum of a
calibrator star observed in the same manner as the target star
immediately before or after the target star.  Figure 2 shows an
azimuthally-averaged power spectrum of Altair and the corresponding
sky power spectrum, compared with a power spectrum of calibrator
Gamma Aquila and its corresponding sky power. 

We then averaged all the calibrated power spectra for a given target. 
If the object and calibrator are both unresolved, the average calibrated
power spectrum should be the power spectrum of the delta function: a
constant.  We found that the pixels along the u and v axes of the power
spectra were often contaminated by noise artifacts from the detector
amplifiers, so we masked them out. 

Figures 3 and 4 show the calibrated azimuthally-averaged power spectra
for our target stars.  To compare different power spectra from
the same target, we normalized each
azimuthally-averaged power spectrum so that the geometric mean
of the first 10 data points in each spectrum equals 1. 
For Altair and 61 Cygni A and B we had more than three pairs of
target and calibrator observations, i.e. calibrated power 
spectra, so we show the average of all the spectra and  
error bars representing the 68\% confidence interval for each datum,
estimated from the variation among the individual power spectra.  The
error is primarily due to differences in the
atmosphere-telescope transfer function between object and calibrator.   
None of the calibrated power spectra deviate from a straight line by more
than a typical error; all the targets are unresolved to the
accuracy of a our measurements.

\section{DISCUSSION}

To interpret our observations we compared them to models of the
IR emission from the solar zodiacal cloud.
We constructed a model for exo-zodiacal emission based on the smooth
component of the Kelsall et al.
\markcite{kels98}(1998) model of the solar system zodiacal cloud as
seen by COBE/DIRBE, with emissivity $\epsilon \propto r^{-0.34}$ and a
temperature  $T = 286 \ {\rm K} \ r^{-0.467}L^{0.234}$,
where $r$ is the distance from the star in AU, and $L$ is the
luminosity of the star in terms of $L_\odot$. 
For a dust cloud consisting entirely of a single kind of dust
particle of a given size and albedo, the $L$ exponent in the
expression for the temperature is simply $-1/2$ times the $r$ exponent
\markcite{back93}(Backman \& Paresce 1993).

The physics of the innermost
part of the solar zodiacal dust is complicated
(see \markcite{mann93} Mann \& MacQueen 1993), but our results are not
sensitive to the details,
because the hottest dust is too close to the star for us to resolve. 
We assume that the dust sublimates at a temperature
of 1500 K, and allow this assumption to define the inner radius of the
disk.  We set the outer radius of the
model to 3 AU, the heliocentric distance of the
inner edge of our own main asteroid belt.  
Our conclusions are not sensitive to this assumption;
decreasing the outer radius to 2 AU or increasing it to
infinity makes a negligible difference in the visibility of the
model, even for A stars.

The assumed surface density profile, however, does make a difference.
A collisionless cloud of dust in approximately circular orbits spiraling into a
star due to Poynting-Robertson drag that is steadily replenished at its outer
edge attains an equilibrium surface density
that is independent of radius \markcite{wyat50}\markcite{brig62}
(Wyatt and Whipple 1950, Briggs 1962). Models that fit data
from the Helios space probes \markcite{lein81}(Leinert et al
1981), the fit by Kelsall et al. \markcite{kels98}(1998) to the
COBE/DIRBE measurements and Good's \markcite{good97}(1997) revised
fit to the IRAS data all have surface densities
that go roughly as $r^{-0.4}$.  This distribution appears to
continue all the way in to the solar corona \markcite{macq95}(MacQueen
\& Greely 1995).  We find that in general, if we assume an
$r^{-\alpha}$ surface density profile, our upper limit for the
1 AU density of a given disk scales roughly as $10^{\alpha/2}$; disks
with more dust towards the outer edge of the 11.6 micron emitting region
are easier to resolve.

Likewise, the assumed temperature profile strongly affects our upper
limits.  Unfortunately, we know little about the temperature profile of the
solar zodiacal cloud.  COBE/DIRBE and IRAS only probed the dust
thermal emission near 1 AU, and Helios measured the solar system cloud in
scattered light, which does not indicate the dust temperature.
We found that a dust cloud model with the IRAS temperature profile
($T = 266 \ {\rm K} \ r^{-0.359}L^{0.180}$) was much easier to resolve
than the model based on DIRBE measurements that we present here,
especially for G and K stars.

To compare the models with the observations, we synthesized high
resolution images of the model disks
at an inclination of 30 degrees.  We calculated the
IR flux of the stars from the blackbody function, and obtained the parallaxes
of the stars from the Hipparcos Catalog \markcite{esa97}(ESA 1997).   We
inferred stellar radii and effective temperatures for each
star from the literature and checked them by comparing the blackbody
fluxes to spectral energy distributions based on photometry from
the SIMBAD database \markcite{simbad} (Egret et al. 1991).  For
Altair and Vega, we use the interferometrically measured angular diameters
\markcite{hanb74}(1974) (they are 2.98 +/-0.14 mas and 3.24 mas).
Stellar fluxes typically disagree with fitted blackbody curves by $\sim10\%$ in the mid-infrared \markcite{enge90}(Engelke 1990), but our method does not require precise photometry, and the blackbody numbers
suffice for determining conservative upper limits.
We computed the power spectra of the images, and normalized them just like
the observed power spectra.  In figures 3 and 4, the azimuthally-averaged
power spectra for our target stars are compared to the extrapolated
COBE/DIRBE model at a range of model surface densities.  Disks 
with masses as high as $10^3$ times the mass of the solar disk will suffer
collisional depletion in their inner regions, so they are unlikely to have
the same structure as the solar disk.  By neglecting this effect we are
being conservative in our mass limits.  The density of the densest model
disk consistent with the data in each case is listed in table 1.

Altair

Our best upper limit is for Altair (spectral type A7, distance 5.1 pc);
with 11 pairs of object and calibrator observations
we were able to rule out a solar-type disk a few
times $10^3$ as dense as our zodiacal cloud.  Such a disk would have been
marginally detectable by IRAS as a photometric excess. 

Vega

IRAS detected no infrared excess in Vega's spectral energy distribution
at 12 microns, with an uncertainty of 0.8 Jy. This may be
due to a central void in the disk interior to about 26
AU \markcite{back93} (Backman \& Paresce 1993).  \markcite{auma84}Aumann et al
(1984) suggested that Vega (A0, 7.8 pc) could have a hot grain component (500 K)
with up to $10^{-3}$ of the grain area of the observed component and not
violate this limit.  The apparent upward trend in the visibility data may
be a symptom of resolved flux in the calibrator stars.  We have only 3 object/calibrator pairs for Vega, not enough to test this hypothesis.
Our upper limit is a solar-type disk with approximately $3 \times 10^3$
times the density of the solar disk.  This disk would have a $\geq$500 K
emitting area of $10^{24} \ {\rm cm}^2$, about $10^{-3}$ of the grain area
of the observed component. 

61 Cygni A and B

Though 61 Cygni is close to the galactic plane and
surrounded by cool cirrus emission, \markcite{back86}Backman, Gillett and Low (1986)
identified an IRAS point source with this binary system and deduced a
far-infrared excess not unlike Vega's.  The color temperature of the excess suggests the presence of dust at distances $> 15$ AU from either star.
However these stars are dim (spectral types
K5 and K7) and the region of the disk hot enough to emit strongly
at 11.6 microns is close to the star and difficult to resolve; we
could not detect a solar-type dust disk around either of these
objects at any density, assuming the COBE/DIRBE model, or unless it
had $10^{5}$ times the density of the solar disk, assuming the IRAS model. 

70 Oph B

70 Oph is a binary (types K0 and K4) with a separation of 24 pixels (2.6 arcsec).   We were able to assemble a power spectrum for B from 9 object/calibrator pairs, but the image of A fell on a part of the LWS chip that suffered from many bad pixels and was unusable.  The image of A may also have been distorted by off-axis effects.  70 Oph B, like 61 Cygni A and B, is dim, making any dust around it cool and hard to detect at 11.6 microns.

$\tau$ Ceti

IRAS could have barely detected a disk with $\sim 1000$ times the emitting area of the solar disk around Tau Ceti (G8, 3.6 pc),
the nearest G star. We have only three object/calibrator pairs for
this object, not enough data to improve on this limit. 

\acknowledgments

  We are grateful to Dana Backman, Alycia Weinberger, Keith Matthews and Eric Gaidos for helpful discussions, and to Keith Matthews and Shri Kulkarni for assistance with the observations.  This research has made use of the Simbad database, operated at CDS, Strasbourg, France.  The observations
reported here were obtained at the W. M. Keck Observatory, which
is operated by the California Association for
Research in Astronomy, a scientific partnership among California Institute of Technology, the University of California, and the National Aeronautics
and Space Administration.  It was made possible by the generous financial support of the W. M. Keck Foundation.

\newpage

\begin{deluxetable}{ccccccc}
\tablenum{1}
\tablewidth{0pt}
\scriptsize
\tablecaption{\label{tbl-1}}
\tablecaption{Observations}
\tablehead{
 & & & \colhead{Exposure Time}& \colhead{Object} & \colhead{} & \colhead{Log Disk Density} \nl
\colhead{Date} & \colhead{Target} & \colhead{Calibrator}   & \colhead{Per Frame (ms)}   & \colhead{Frames}  & \colhead{Pairs} &
\colhead{Solar Disk = 0\tablenotemark{a}}    
} 
\startdata
August 3 & Vega& R Lyr& 800 & 90& 2 & $\le 4.0$\nl
 & &$\kappa$ Lyr&800 &90& 1 &\nl
 & 61 Cyg A & $\zeta$ Cyg & 800 & 90 & 4 & \dots \nl
 & 61 Cyg B & $\zeta$ Cyg & 800 & 90 & 5 & \dots \nl
 &$\tau$ Cet & $\nu$ Cet & 800 & 90 &2 & \dots \nl
\medskip & &  & 210 & 348&  1  &\nl
August 4 & 70 Oph B& $\beta$ Oph  &84 &864 &  5 &\dots \nl
  & &  74 Oph & 84 & 864 & 4 & \nl
  & Altair &  $\gamma$ Aql& 84 & 864 & 6 &$\le 3.2$\nl
 & & $\beta$ Aql &84  &864 & 5 &\nl
\enddata

\tablenotetext{a}{based on the COBE/DIRBE model of the solar zodiacal cloud (Kelsall et al. 1998)}

\end{deluxetable}

\newpage

\figcaption{ A cut through a single 4 ms image of Altair, compared to a similar cut through an image of a calibrator star, Gamma Aquila, and an Airy function representing the PSF of an ideal, filled, 10-meter aperture at 11.6 microns.  The cores of the images are diffraction-limited, but the wings are sensitive to the instantaneous seeing, making speckle analysis necessary. \label{fig1}}

\figcaption{ An azimuthally-averaged power spectrum of Altair and the corresponding sky power spectrum, compared with a power spectrum of calibrator
Gamma Aquila and its corresponding sky power.  The power in the star images approaches the sky power near the diffraction limit at 4 cycles per arcsecond.  \label{fig2}}

\figcaption{ Azimuthally-integrated power spectrum of Altair compared to simulated power spectra of model disks with various densities (1 = the solar disk).  An unresolved point-source would appear as a straight line at a normalized power of 1.0. The densest model disk consistent with the observations has a density of roughly $10^3$ times that of the solar disk\label{fig3}}

\figcaption{ Azimuthally-integrated power spectra of other target stars compared to simulated power spectra of model disks with various densities. \label{fig4}}

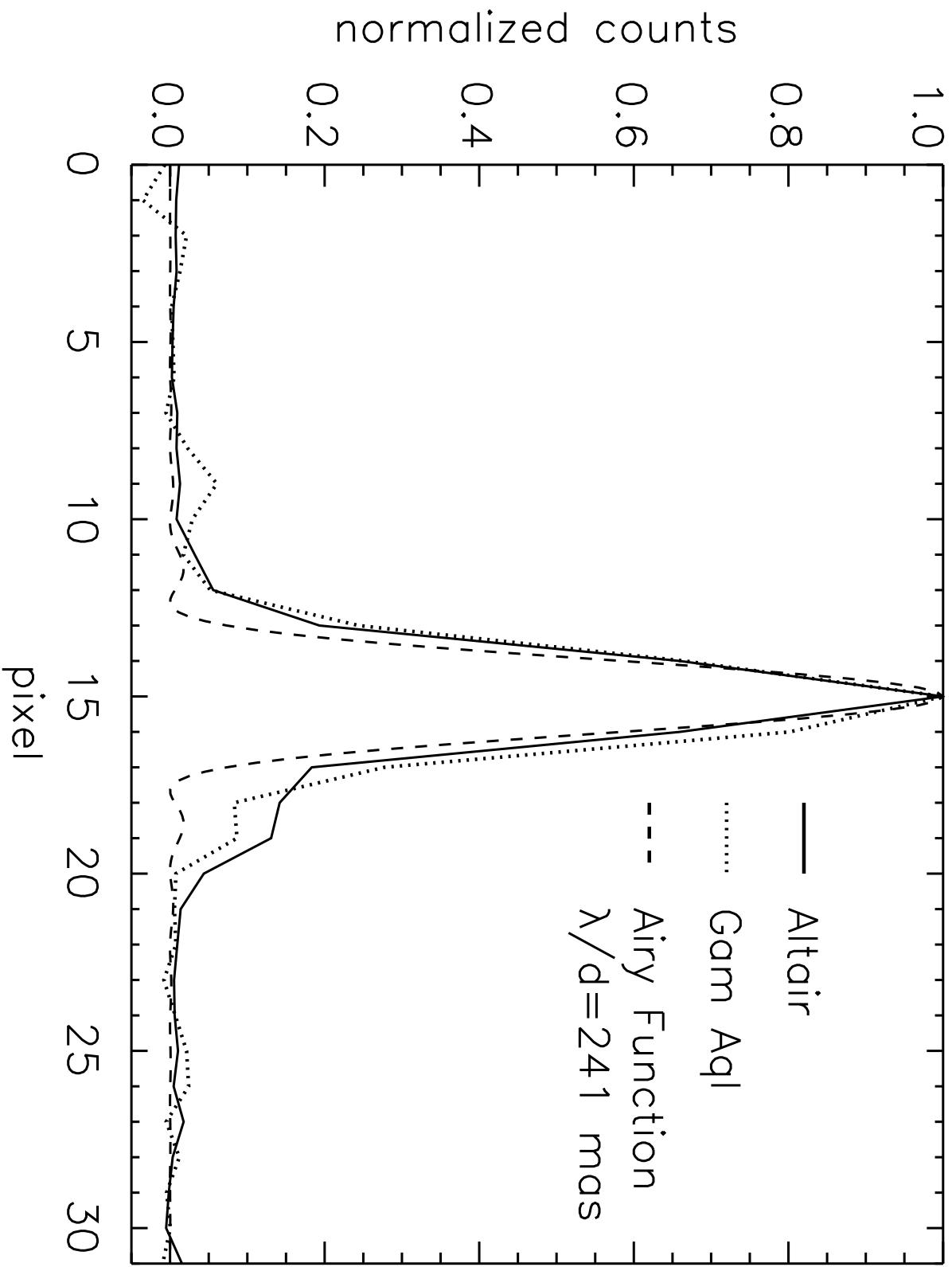
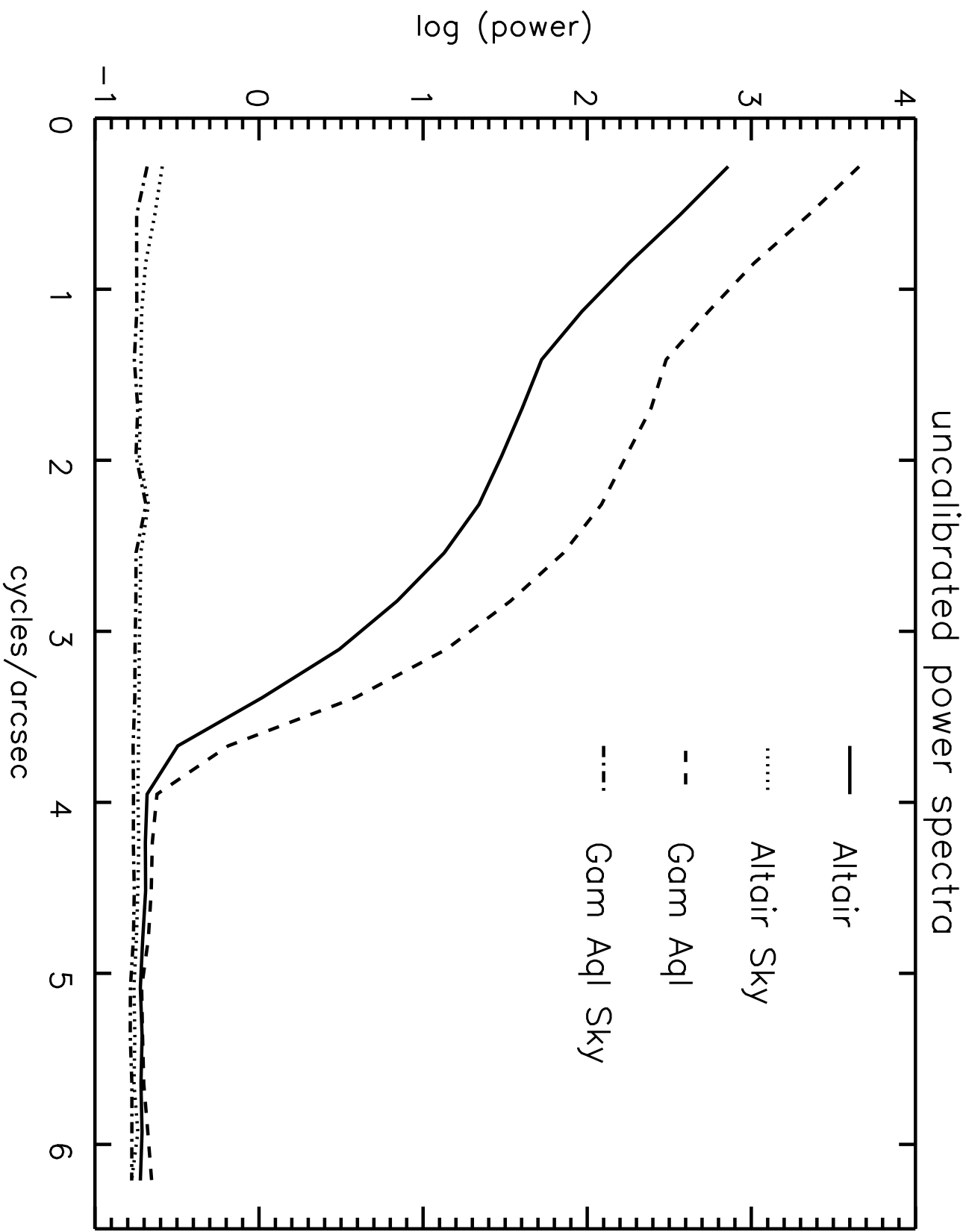
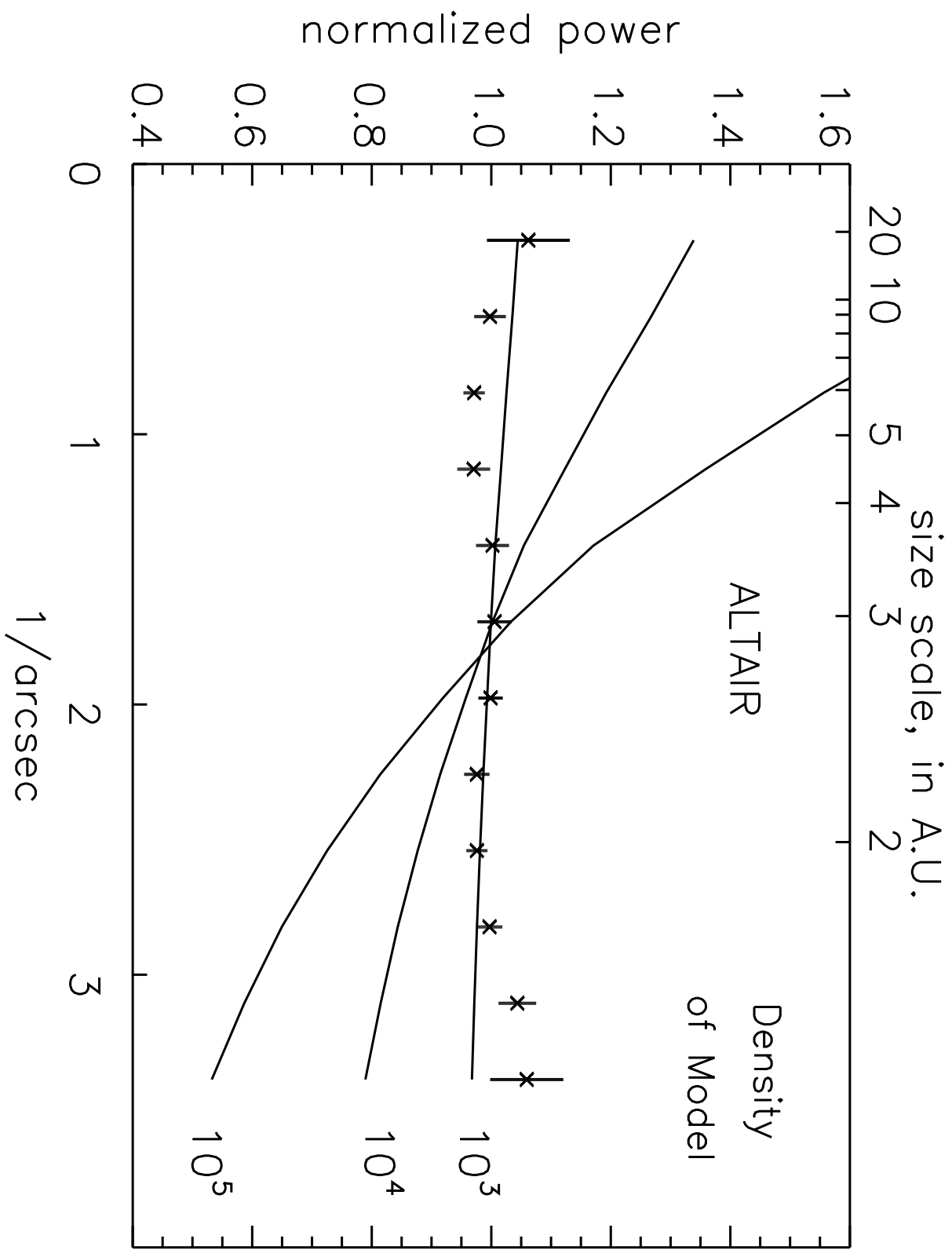
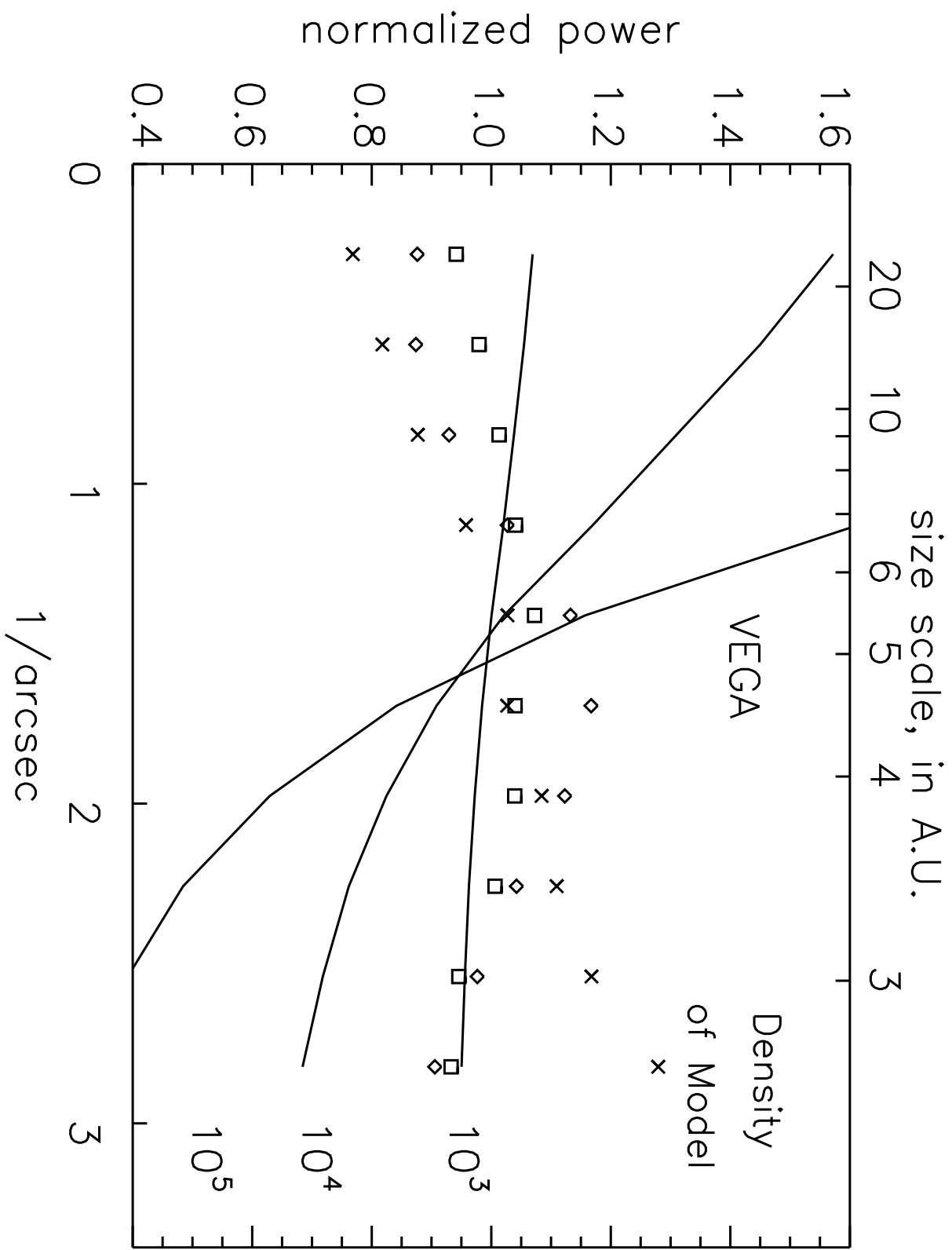
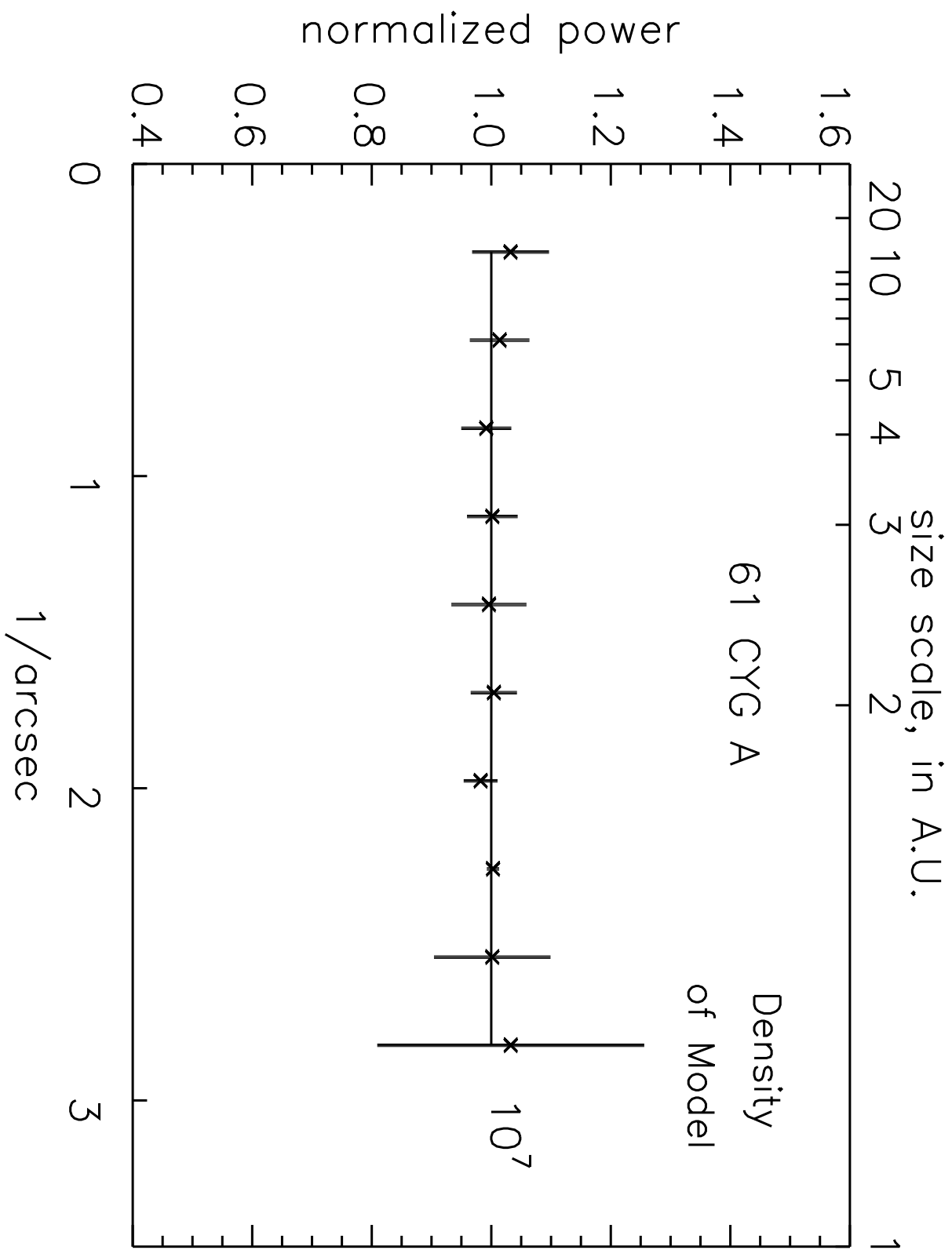
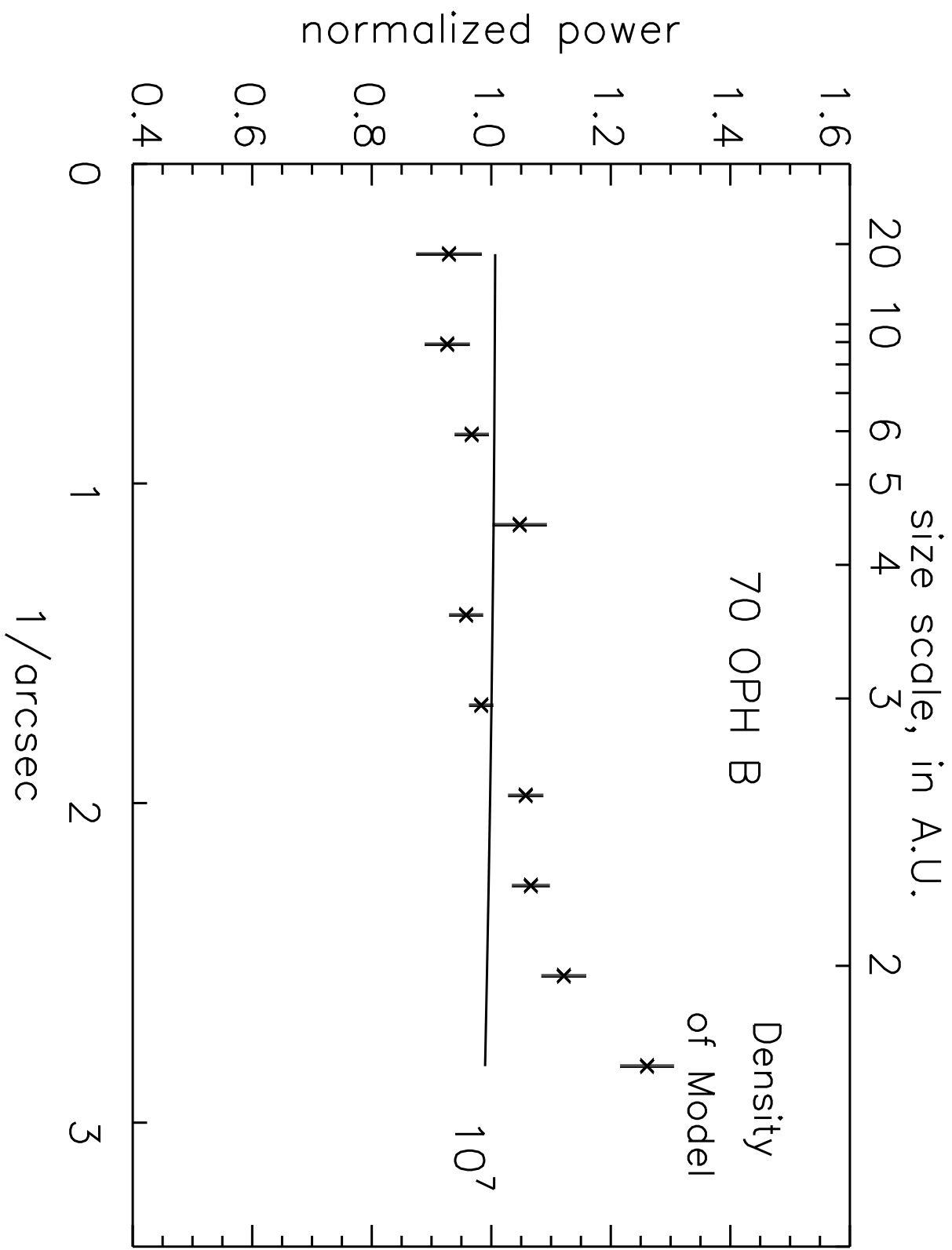
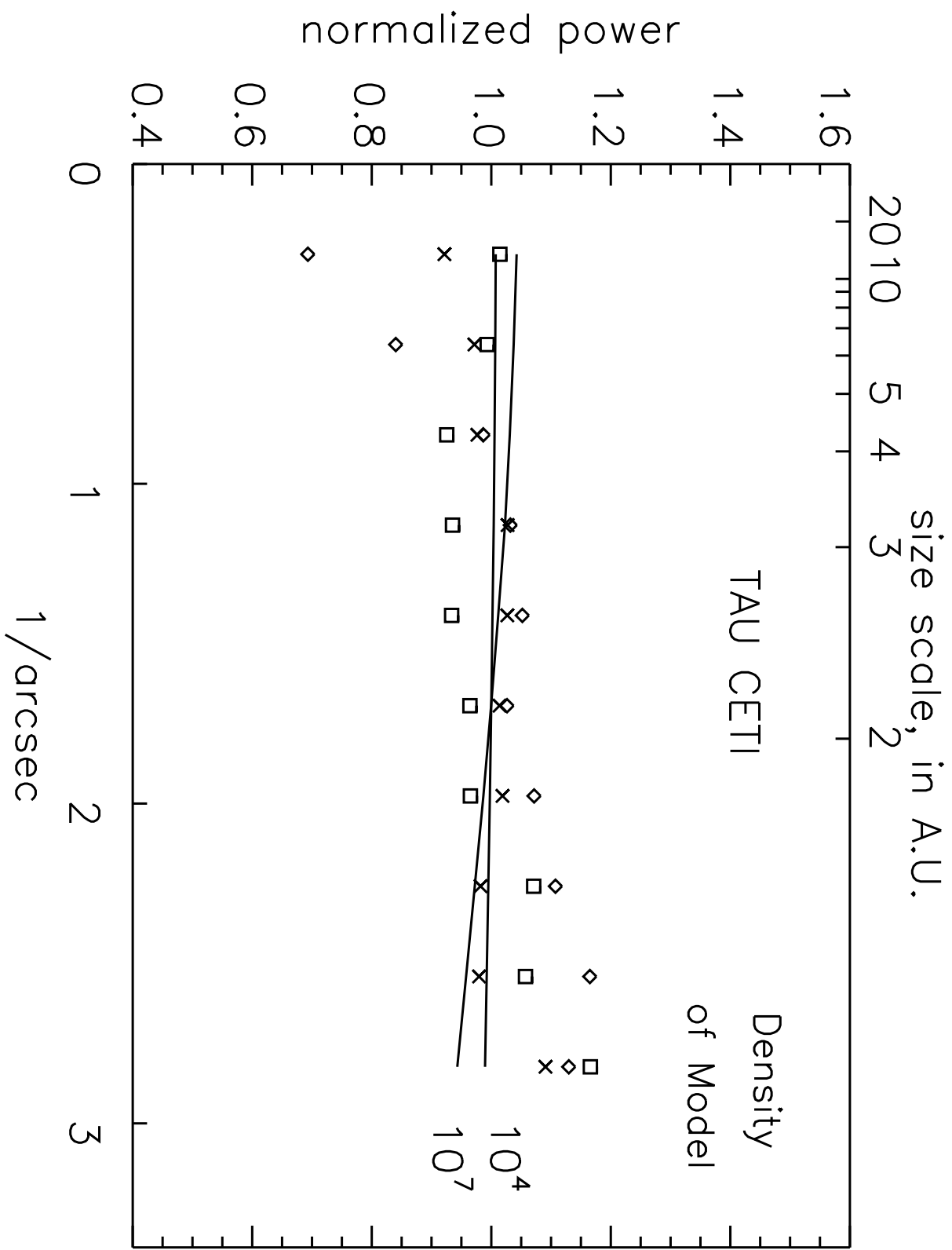
\end{document}